\documentclass[english]{revtex4}
\usepackage[T1]{fontenc}
\usepackage[latin1]{inputenc}
\usepackage{graphicx}

\makeatletter

\providecommand{\tabularnewline}{\\}

\usepackage{epsfig}

\usepackage{babel}
\makeatother
\begin{document}
\def\bivec#1{\vbox{\ialign{##\crcr$\leftrightarrow$\crcr\noalign{\kern-1pt\nointerlineskip}$\hfil\displaystyle{#1}\hfil$\crcr}}}

\title{Parallel multiscale modeling of biopolymer dynamics with hydrodynamic
correlations}

\author{Maria Fyta$^{(a)}$}

\author{Jayanta Sircar$^{(b)}$}

\author{Efthimios Kaxiras$^{(a,b)}$}

\affiliation{$^{(a)}$Department of Physics and $^{(b)}$School of Engineering and Applied Sciences,
Harvard University, Cambridge MA 02138, USA }

\author{Simone Melchionna}

\affiliation{SOFT-INFM-CNR and Department of Physics, University of Rome \textit{La
Sapienza,} P.le A. Moro 2, 00185, Rome, Italy}

\author{Massimo Bernaschi}

\author{Sauro Succi}

\affiliation{Istituto Applicazioni Calcolo, CNR, Viale del Policlinico 137, 00161,
Rome, Italy }

\begin{abstract}
We employ a multiscale approach to model the translocation of biopolymers
through nanometer size pores. Our computational scheme combines microscopic
Molecular Dynamics (MD) with a mesoscopic Lattice Boltzmann (LB) method
for the solvent dynamics, explicitly taking into account the interactions
of the molecule with the surrounding fluid. We describe an efficient
parallel implementation of the method which exhibits excellent scalability
on the Blue Gene platform. We investigate both dynamical and statistical aspects
of the translocation process by simulating polymers
of various initial configurations and lengths. For a representative
molecule size, we explore the effects of important parameters that
enter in the simulation, paying particular attention to the strength
of the molecule-solvent coupling and of the external electric field
which drives the translocation process. Finally, we explore the connection
between the generic polymers modeled in the simulation and DNA, for
which interesting recent experimental results are available. 
\end{abstract}
\maketitle

\section{Introduction \label{sec_intro} }

Biological systems exhibit a complexity and diversity much richer
than the simple solid or fluid systems traditionally studied in physics
or chemistry. The powerful quantitative methods developed in the latter
two disciplines to analyze the behavior of prototypical simple systems
are often difficult to extend to the domain of biological systems.
Advances in computer technology and breakthroughs in simulational
methods have been constantly reducing the gap between quantitative
models and actual biological behavior. The main challenge remains
the wide and disparate range of spatio-temporal scales involved in
the dynamical evolution of complex biological systems. In response
to this challenge, various strategies have been developed recently,
which are in general referred to as ``multiscale modeling''. These
methods are based on composite computational schemes in which information
is exchanged between the scales, in either a sequential or a concurrent manner
\cite{MMS_review}.

We have recently developed a multiscale framework which is well suited
to address a class of biologically related problems. This method involves
different levels of the statistical description of matter (continuum
and particle) and is able to handle different scales through the
spatial and temporal coupling of a \textit{mesoscopic} fluid solvent,
based on the lattice Boltzmann method \cite{LBEa,LBEb,LBEc} (LB), 
to a coarse-grained particle
level, which employs explicit molecular dynamics (MD). The solvent
dynamics does not require any form of statistical ensemble averaging
as it is represented through a discrete set of pre-averaged probability
distribution functions, which are propagated along straight particle
trajectories. This dual field/particle nature greatly facilitates
the coupling between the mesoscopic fluid and the atomistic level,
which proceeds seamlessy in time and only requires standard 
interpolation/extrapolation methods
for information-transfer in physical space. Full details on this scheme
are reported in Ref. \cite{ourLBM}. 
LB and MD with Langevin dynamics have been
coupled before \cite{DUN}, but our implementation 
involves such coupling for long molecules of biological interest. In addition,
we have recently developed a parallel version of the code and successfully
ported it to the IBM-BlueGene architecture. 

Motivated by recent experimental studies, we apply this multiscale
approach to the translocation of a biopolymer through a narrow pore.
This type of biophysical process is important in phenomena like
viral infection by phages, inter-bacterial DNA transduction or gene
therapy \cite{TRANSL}. In addition, it is hoped that this type of process will open a way
for ultrafast DNA-sequencing by sensing the base-sensitive electronic  
signal as the biopolymer passes through a nanopore with attached electrodes. 
Experimentally, translocation is observed
\textit{in vitro} by pulling DNA molecules through micro-fabricated
solid state or membrane channels under the effect of a localized electric
field \cite{EXPRM}. From a theoretical point of view, simplified
schemes \cite{statisTrans} and non-hydrodynamic coarse-grained or
microscopic models \cite{DynamPRL,Nelson} are able to analyze universal
features of the translocation process. This, though, is a complex
phenomenon involving the competition between many-body interactions
at the atomic or molecular scale, fluid-atom hydrodynamic coupling,
as well as the interaction of the biopolymer with wall molecules in
the region of the pore. A quantitative description of this complex
phenomenon calls for state-of-the art modeling, towards which the
results presented here are directed.

\section{Coupling between Lattice Boltzmann and Molecular Dynamics}

In this section we outline the simulation method used to couple the
Lattice Boltzmann description for the solvent to the Molecular Dynamics description
of the solute. 
In the LB method the basic quantity is $f_{i}(\vec{x},t)$, representing
the probability of finding a ``fluid particle'' at the spatial mesh location
$\vec{x}$ and at time $t$ with discrete speed $\vec{c}_{i}$. 
We emphasize that the ``fluid particles'' do not correspond to individual 
physical particles such as water molecules; they are simply an effective medium 
for representing the collective motion of such physical particles.  Once
the discrete distributions $f_{i}$ are known, the local density,
flow speed and momentum-flux tensor of the fluid are obtained by a direct summation
upon all discrete distributions: 
\begin{equation}
\rho(\vec{x},t)=\sum_{i}f_{i}(\vec{x},t)\label{dens}\end{equation}
 \begin{equation}
\rho\vec{u}(\vec{x},t)=\sum_{i}f_{i}(\vec{x},t)\vec{c}_{i}\label{vel}\end{equation}
 \begin{equation}
\bivec P(\vec{x},t)=\sum_{i}f_{i}(\vec{x},t)\vec{c}_{i}\vec{c}_{i}\label{pre}\end{equation}
where, for the present study, the standard three-dimensional 19-speed
lattice is used \cite{LBEc}. The fluid populations are advanced in
time through the following evolution equation: 
\begin{equation}
f_{i}(\vec{x}+\vec{c}_{i}\Delta t,t+\Delta t)=f_{i}(\vec{x},t)+\omega\Delta t(f_{i}-f_{i}^{eq})(\vec{x},t)+F_{i}\Delta t+S_{i}\Delta t\label{lbe2}
\end{equation}
The discrete velocities $\vec{c}_{i}$ connect mesh points to first and
second topological neighbors, therefore the fluid particles
can only move along the links of a regular lattice and the synchronous
particle displacements $\Delta\vec{x}_{i}=\vec{c}_{i}\Delta t$ never
take the fluid particles away from the lattice. The right hand side
of Eq. (\ref{lbe2}) represents the effect of fluid-fluid molecular
collisions, through a relaxation towards a local equilibrium, $f_{i}^{eq}$,
typically a second-order expansion in the fluid velocity of a local
Maxwellian with speed $\vec{u}$: 
\begin{equation}
f_{i}^{eq}=w_{i}[\beta\vec{u}\cdot\vec{c}_{i}+
\frac{\beta^{2}}{2}(\vec{u}\vec{u}\cdot(\vec{c}_{i}\vec{c}_{i}-\frac{1}{\beta}\bivec I)]
\end{equation}
where $\beta=1/k_{B}T$ is the inverse temperature, $w_{i}$ a set
of weights normalized to unity, and $\bivec I$ is the unit tensor
in cartesian space. The relaxation frequency $\omega$ controls the
kinematic viscosity of the pure fluid: \[
\nu=k_{B}T\left(\frac{1}{\omega}-\frac{\Delta t}{2}\right)\]

The $F_{i}$ term is a stochastic force needed to inject fluctuations
in the fluid at the level of fluctuating hydrodynamics, which is 
local in space and time, and is given by
\begin{equation}
F_{i}=w_{i}\{
\sum_{a,b} F^{(2)}_{ab} (c_{ia}c_{ib} - k_BT \delta_{ab}) +
\sum_{a,b,c} F^{(3)}_{abc} g_{iabc}
\}
\end{equation}
where $F^{(2)}$ is a fluctuating stress tensor satisfying
\begin{equation}
\langle 
 F^{(2)}_{ab}(\vec{x},t) F^{(2)}_{cd} (\vec{x}',t')\rangle = 
\frac{\gamma k_B T}{m} \Delta_{abcd} \delta(\vec{x}-\vec{x}')\delta(t-t')
\end{equation}
and $\Delta_{abcd}$ is a fourth-order Kronecker symbol.
By construction, the fluctuating stress tensor does not affect the fluid mass and momentum 
conservation. Moreover, due to the discrete nature of the LB scheme, and in order to recover
the fluctuation-dissipation theorem at finite wave vectors, noise also
acts at the level of the non-hydrodynamic modes carried by the
LB method via a fluctuating heat flux $F^{(3)}$, coupling through a suitable
basis in kinetic space, $g$ (see ref. \cite{adhikari} for full details).

The source term $S_{i}$ accounts for the presence of atomic-scale particles embedded
in the LB solvent; we use here the term particles again, because in the 
general case those may represent individual atoms or collections of atoms (molecular
units), or even coarse-grained descriptions of atomic motion, but still
at the atomic length scale (a few \AA~ to a few nm). 
$S_{i}$ is a back-reaction representing the sum of
momentum and momentum-flux input per unit time due to the influence
of atomic-scale particles on the fluid population $f_{i}$. By definition, the back-reaction
does not change the density of the fluid, so that $\sum_{i}S_{i}=0$.
Momentum and momentum flux conservation of the solute and solvent systems
imply that 
\begin{equation}
\sum_{i}S_{i}(\vec{x})\vec{c}_{i}=-(\vec{F}^{f}(\vec{x})+\vec{F}^{r}(\vec{x}))
\label{momentum_cons}
\end{equation}
The source term then reads
\begin{equation}
S_{i}=-w_{i}\beta[\vec{F}^{f}(\vec{x})+\vec{F}^{r}(\vec{x})]\cdot\vec{c}_{i}
\end{equation}
so that it explicitly satisfies Eq. (\ref{momentum_cons}).
In these equations, $\vec{F}^{f}$ and $\vec{F}^{r}$, are the friction and 
random forces, respectively.

Before describing the MD part, we emphasize that the LB solver is
particularly well suited to the problem at hand for several reasons: First, free-streaming
of the fluid proceeds along straight trajectories which greatly facilitates the
imposition of geometrically complex boundary conditions, such as those 
required to describe the membrane and nano-pore. Second,
fluid diffusivity emerges from the first-order LB relaxation-propagation
dynamics, so that the kinetic scheme can march in time-steps which
scale only linearly with the mesh resolution. Third, since collisions
are completely local, the LB scheme is ideally suited to parallel
computing. These features make the LB the method of choice compared to
other available methods, such as Stokesian dynamics, which typically
scale superlinearly with the number of particles.

We next describe the MD section of the method, bearing in mind that
the embedded solute has a molecular topology, such as DNA, where a
linear collection of $N_0$ beads (each bead or solute particle representing 
a collection of atoms or molecules) compose the polymer. The solute particles
are advanced in time according to the following MD
equations for the positions $\vec{r}_{p}$ and velocities $\vec{v}_{p}$
\begin{eqnarray}
\frac{d\vec{r}_{p}}{dt} & = & \vec{v}_{p}\nonumber \\
m\frac{d\vec{v}_{p}}{dt} & = & \vec{F_{p}}^{c}+\vec{F_{p}}^{f}+\vec{F_{p}}^{r}+\vec{F_{p}}^{b},\;\;\;\;\; p=1,N_0\label{MD}
\end{eqnarray}
 where the forces on the right-hand side are given by 
\begin{eqnarray}
\vec{F_{p}}^{c} & = & -\sum_{q}\partial_{\vec{r}_{p}}V(|\vec{r}_{p}-\vec{r}_{q}|)\label{MDFORCES}\\
\vec{F_{p}}^{f} & = & \gamma(\vec{u}_{p}-\vec{v}_{p})\\
\vec{F_{p}}^{r} & = & \vec{\xi}(\vec{r}_{p},t)
\end{eqnarray}
The first term represents the conservative particle-particle interactions,
$V(r)$ being a standard $6-12$ Lennard-Jones potential,
\begin{equation}
V(r)=4\epsilon\left[\left(\frac{\sigma}{r}\right)^{12}-\left(\frac{\sigma}{r}\right)^{6}\right] 
\end{equation}
with an effective
cut-off at $r_{c}=2^{1/6}\sigma$ for the radial part, plus a harmonic
potential for angular degrees of freedom, 
$V_{ang}(\phi)=\frac{\kappa_{ang}\phi^{2}}{2}$,
with $\phi$ the relative angle between two consecutive bonds, to
account for distortions of the dihedral angles. 
 Typical values of $\sigma$ and $\epsilon$ in our simulations are
1.8 and $10^{-4}$, respectively.
The
second term on the right-hand-side of (\ref{MD}) represents the mechanical
friction between a single particle and the surrounding fluid, $\vec{v}_{p}$
being the particle velocity and $\vec{u}_{p}\equiv\vec{u}(\vec{r}_{p})$
the fluid velocity, evaluated at the particle position. In addition to
the mechanical drag, the particles feel the effects of stochastic fluctuations
of the fluid environment through the random term $\vec{\xi}(\vec{r}_{p},t)$,
which is a Gaussian random noise obeying the fluctuation-dissipation
relations, \begin{eqnarray}
<\vec{\xi}(\vec{r}_{p},t)> & = & 0\nonumber \\
<\vec{\xi}(\vec{r}_{p},t)\vec{\xi}(\vec{r}_{q},t')> & = & \frac{2\gamma k_{B}T}{m}\delta(\vec{r}_{p}-\vec{r}_{q})\delta(t-t')\end{eqnarray}
Finally, $\vec{F_{p}}^{b}$ corresponds to the bonding forces acting
between particles with labels $p$ and $p+1$ of the polymeric chain. The bonding
forces can be modelled as arising from a rigid constraint that fixes
the bond length to a constant value $r_{0}$. In this case, 
\begin{equation}
\vec{F_{p}}^{b}=\sum_{k}\lambda_{k}\partial_{\vec{r}_{p}}\sigma_{k}
\end{equation}
is the reaction force resulting from the constraint 
\begin{eqnarray}
\sigma_{k} & \equiv & |\vec{r}_{k+1}-\vec{r}_{k}|^{2}-r_{0}^{2}=0\label{sigma}
\end{eqnarray}
and $\{\lambda\}$ is the set of Lagrange multipliers ($N_0-1$ in the
case of a linear polymer of length $N_0$) that depend instantaneously on the particle
positions and momenta. The Lagrange multipliers are evaluated based
on the numerical scheme used to propagate in time the equations of
motion. The linear dependence of the dynamics from the Lagrange multipliers
requires the inversion of a linear matrix which depends on the position
and momenta of all particles composing the polymer. Such direct inversion
can be avoided through the well-known SHAKE method \cite{SHAKE}, an iterative
procedure which solves the algebraic problem to a given accuracy.
The SHAKE method, however, becomes rather impractical in a parallel architecture,
since at each iteration it requires frequent exchange of data, therefore
representing a bottleneck in terms of scalability.

As an alternative, which is particularly well suited for the parallel implementation,
bonding can be modelled by harmonic forces:
\begin{eqnarray}
\vec{F}_{b}^{b}&=&-\partial_{\vec{r}_{p}}V^{b}
\\
V^{b}&=&\sum_{k}\frac{\kappa_r}{2}[|\vec{r}_{k+1}-\vec{r}_{k}|-r_{0}]^{2}
\end{eqnarray}
In order to reduce the additional polymer flexibility due to such
forces, a rather high value of the force constant $\kappa_r$ can be chosen.
The basic difference with the constrained dynamics is the fact that
harmonic bonding introduces fast oscillations which can render the
numerical scheme unstable at large timesteps. Typically, such modes
carry frequencies up to two orders of magnitude higher than those
relative to non-bonding forces. To take into account such oscillations,
a small integration time step must be used which would make the
simulation highly inefficient. On the other hand, as described in
the following, a multiple time step algorithm makes it possible to achieve basically
the same computational efficiency for the constrained and unconstrained
MD schemes. 

Clearly, in the LB approach all quantities have to reside on the lattice
nodes, which means that the frictional and random forces need to be
extrapolated from the particle to the grid location. This is obtained
by extracting the fluid velocity field $\vec{u}_{p}$ at the nearest
grid point from each particle position and, similarly, assigning these
forces to the feed-back on the fluid population through the same simple
recipe. We have found that this procedure is as accurate as a more
involved bilinear interpolation/extrapolation scheme for the exchange
of forces and momentum. 
 Details on this scheme have been published previously (see Ref. \cite{ourLBM}).

The numerical solution of the stochastic equations is achieved through
the Stochastic Position Verlet (SPV) scheme, as introduced in Ref.
\cite{SMJCP}, a propagator which is second order accurate in time. 
Owing to the presence of velocity-dependent and stochastic
forces, standard deterministic integrators, such as the Verlet one, would give
rise to first order accuracy of the resulting trajectory.
The original
SPV method needs to be modified in the presence of constraining forces.
Positions and momenta are advanced in time according to 
\begin{eqnarray}
\vec{r}_{p} & = & \vec{r}_{p}+\frac{dt}{2}\vec{v}_{p} \nonumber \\
\vec{r}_{p} & \rightarrow & \vec{r}_{p}^{\sigma}\mbox{ (SHAKE) } \nonumber \\
\vec{v}_{p} & = & e^{-\gamma dt}\vec{v}_{p}+\frac{1-e^{-\gamma dt}}{m\gamma}\vec{F}_{p}(\vec{r}^{\sigma})+{\cal N}[\frac{k_{B}T}{m}(1-e^{-2\gamma dt})] \nonumber \\
\vec{v}_{p} & \rightarrow & \vec{v}_{p}^{\sigma}\mbox{ (RATTLE) } \nonumber \\
\vec{r}_{p} & = & \vec{r}_{p}^{\sigma}+\frac{dt}{2}\vec{v}_{p}^{\sigma} \label{SPValgo}
\end{eqnarray}
where ${\cal N}[\sigma^{2}]$ is a gaussian random number of zero mean and
variance $\sigma^{2}$ , and
\begin{eqnarray}
\vec{F}_{p}=\vec{F}_{p}^{c}+\gamma\vec{u}_{p}+\vec{F}_{p}^{r}
\end{eqnarray}
and finally, the SHAKE procedure \cite{SHAKE} is employed to satisfy
constraints, $\{\sigma(\vec{r})\}=0$, while the RATTLE procedure
\cite{RATTLE} imposes the set of conditions 
$\{\dot{\sigma}(\vec{r},\vec{v})\}=0$.
Clearly, the computational effort of the SPV integrator is the same as
in standard MD, where the conventional Verlet algorithm is usually employed, 
by computing only once and storing the exponential factors appearing 
in eqs. (\ref{SPValgo}).
When considering the momentum exchange
with the solvent, the corrected velocities appear in the friction
forces. Moreover, during the MD sub-cycle, the hydrodynamic field is frozen
at time $t=n\Delta t$. 

For the translocating polymer, the MD solver is marched in time with
a fraction of the LB time-step, $dt=\Delta t/M$ and the timestep
ratio $M$ is chosen to be $5$.

When dealing with the harmonic bonds a multiple time step integrator
is employed \cite{SMJCP} by introducing a nested sub-cycle over a
timestep $dt^{b}=dt/M^{b}$, as follows: 
\begin{eqnarray}
\vec{r}_{p} & = & \vec{r}_{p}+\frac{dt}{2}\vec{v}_{p}\nonumber \\
\vec{v}_{p} & = & \vec{v}_{p}+\frac{dt}{2m}\vec{F}_{p}(\vec{r})\nonumber \\
 &  & \left\{ \vec{v}_{p}=e^{-\gamma dt^{b}}\vec{v}_{p}+\frac{1-e^{-\gamma dt^{b}}}{m\gamma}\vec{F}_{p}^{b}(\vec{r}^{\sigma})+{\cal N}[\frac{k_{B}T}{m}(1-e^{-2\gamma dt^{b}})]\right\} _{M^{b}\, cycle}\nonumber \\
\vec{v}_{p} & = & \vec{v}_{p}+\frac{dt}{2m}\vec{F}_{p}(\vec{r})\nonumber \\
\vec{r}_{p} & = & \vec{r}_{p}+\frac{dt}{2}\vec{v}_{p}
\end{eqnarray}
The multiple time step solver is now marched in time with timestep
ratios $M=5$ and $M^{b}=20$, providing accurate results in terms
of stability and unbiased statistical averages, as verified by monitoring
the system average temperature which remains equal to the preset value.
More details on the method and the efficiency of each of the schemes involved
(MD and LB) are given in Ref. \cite{ourLBM}.

\section{Code Parallelization}

We have recently developed a
 parallel version of the LB-MD multiscale code
and ported it to the IBM Blue-Gene architecture.  This development provides
a boost in computational efficiency by an order of magnitude in both solvent and solute
degrees of freedom, making possible simulations 
in which one MD particle corresponds to one DNA base-pair.
This level of resolution at the lowest scale of the multiscale approach 
enables the concurrent handling of hydrodynamics with chemical specificity.
In this section, we provide the essential features of the parallel implementation
of the LB-MD code.  

The parallelizations of the Lattice Boltzmann Equation method and 
of the Molecular Dynamics method, separately, 
have been extensively studied for a number of years 
\cite{PARALB,MD1,MD2,MD3,MD4,MD5,MD6}.
However, the coupling of these techniques poses new issues that need
to be addressed in order to achieve scalability and efficiency for large scale
simulations. We addressed these issues starting from a serial version
of the combined code, instead of trying to combine two existing parallel
versions. We chose MPI as the communication interface since it offers
high portability among different platforms and allows good performance
due to the availability of highly tuned implementations. For the
two sections of the code, we employed a parallelization technique
that entails a sort of ``run-time'' pre-processing. For the LB
part of the code, this initial stage can be summarized as follows.
Each node of the LB lattice is labeled with a {\em tag} that identifies
it as belonging to a specific part of the computational domain (
e.g., fluid, wall, boundary, etc.), as read from an input
file. In this way it is possible to use the code for different problems
without recompiling it. At first, a subset of nodes is assigned to
each computational task, attempting to balance the number of nodes
{\em per} task as much as possible (obviously in some cases this
operation cannot be done exactly). The assignment takes into account
the domain decomposition strategy that can be one-, two- or three-dimensional.
All seven possible combinations (that is, decompositions along
$X,Y,Z,XY,XZ,ZY,XYZ$) are supported. The decomposition strategy can
be chosen at run time as well. Having assigned the nodes to the tasks,
the {\em pre-processing} phase begins. Basically, each task globally
checks which tasks own the nodes to be accessed during the subsequent
phases of simulation, for the streaming part of the LB algorithm.
Such information is exchanged by using MPI collective communication
primitives so that each task knows the neighboring peer for send/receive
operations. Information about the type and size of data to be sent/received
is exchanged as well. 

In the LB scheme there are several parts in which data are exchanged:
i) streaming; ii) handling of periodic boundary
conditions; iii) presence of reflecting or absorbing {\em
walls} within the computational domain. For each of these cases we
adopt the following approach: the {\em receive} operations are
posted in advance by using corresponding non-blocking MPI primitives,
then the {\em send} operations are carried out using either blocking
or non-blocking primitives depending on the parallel platform in use
(unfortunately, few platforms allow real overlapping between communication
and computation). Then, each task waits for the completion of its
receive operations using the MPI {\em wait} primitives. The last
operation, in the case of non-blocking {\em send}, is to wait for
their completion. The evaluation of global quantities (e.g.,
the momentum along the $X,Y,Z$ directions) is carried out by
using MPI collective communication primitives of {\em reduction}.
This parallelization scheme works fine for the LB part of the code,
provided that the computational domain remains fixed (that is, each
node maintains the same tag), otherwise a new {\em pre-processing}
step is required.

In absence of particles embedded in the fluid, the performance
of the parallel LB section is very good, as shown in Table \ref{tab:largelbe},
which gives the results for two large lattices obtained using 512 and
1024 processors of the IBM  BlueGene parallel system \cite{bluegene}. Actually,
excellent scalability is achieved even for a much smaller lattice,
as reported in Table \ref{tab:smalllbe}:

\begin{table}[h!]
 \begin{centering}
\begin{tabular}{|c|c|c|}
\hline 
 &
Time (sec) &
Time (sec) \tabularnewline
Lattice size &
512 tasks &
1024 tasks\tabularnewline
\hline  
$512\times256\times256$&
71 &
35\tabularnewline
$1024\times512\times512$&
584 &
286\tabularnewline
\hline
\end{tabular}
\par\end{centering}
\caption{Times (in seconds) for 100 LB time-steps using 512 and 1024 IBM BlueGene processors}

\label{tab:largelbe} 
\end{table}

\begin{table}[h!]

\begin{centering}
\begin{tabular}{|c|c|c|c|}
\hline 
Number of tasks &
Time (sec) &
Number of nodes {\em per} task &
Efficiency  (\%)\tabularnewline
\hline 
32 &
69.6 &
65536 &
\tabularnewline
128 &
17.4 &
16384 &
100 \tabularnewline
512 &
4.38 &
4096 &
99.4 \tabularnewline
1024 &
2.32 &
2048 &
93.8 \tabularnewline
\hline
\end{tabular}
\par\end{centering}

\caption{Times for 100 LB time-steps on a $128\times128\times128$ lattice
using the IBM  BlueGene. 
Note that 32 is the minimum number of tasks with the
system configuration used for this test.
}

\label{tab:smalllbe} 
\end{table}

For the Molecular Dynamics section, a parallelization strategy specific
to the multi-scale method in use must be developed. The main problem,
given the highly non-homogeneous distribution of particles for typical
multi-scale applications, which is particularly true in the case of biopolymer translocation
through a nanopore, is that a strong load imbalance is expected (one
task may have been assigned hundreds of particles whereas a neighboring one
may have no particles at all); therefore, a careful partitioning of the particles
is necessary. The usual optimal choice for MD is a domain decomposition
strategy, where as a first option, the parallel sub-domains coincide
with the ones of the LB scheme. In this way, each computational
task performs sequentially both the LB and MD calculations, avoiding
the communication costs arising from a functional decomposition. Alternatively,
a more MD-specific decomposition, which considers only the regions
of the spatial domain populated by particles, could be a better choice
in terms of stand-alone MD. With this second option, however, the
exchange of momentum among particles and the surrounding fluid becomes
a non-local operation, possibly with long-range point-to-point communications
from the viewpoint of the underlying hardware platform, and a consequently
unacceptable communication cost.

We have thus opted for the first strategy, such that the interaction
with the fluid is completely localized (there is no need to exchange
data among tasks during this stage). However, since this subdivision
gives rise to a strong load imbalance, we resort to a {\em dynamic}
load balancing algorithm, as outlined in the following.

At first, during the {\em pre-processing} step a subset of particles
is assigned to the computational tasks. As the simulation proceeds, particles
migrate from one domain to another and particle coordinates, momenta
and identities are re-allocated among tasks. Non-bonding forces between
intra- and inter-domain pairs of particles, involving the communication
of particle positions between neighboring tasks, are computed. Moreover,
the molecular topology is taken into account by exchanging details
on the molecular connectivity, in order to compute bonding forces locally.
In this respect, the way parallel MD is designed is rather standard
\cite{MD6} and will not be described in detail. As a simple
add-on to the standard procedure, each task carries out the exchange
of momentum with the fluid.

The dynamic load balancing strategy impacts the computation of the
non-bonding forces, representing the most CPU demanding part of MD.
At first, the strategy requires a communication operation between
neighboring tasks that tracks the number of particles assigned
to the neighbors. If a task has a number of particles that exceeds by
a predefined threshold the number of particles assigned to one neighbor,
it sends to that neighbor part of the exceeding particles, so that local
load balancing for the computation of non-bonding forces is achieved.
A set of {\em precedence} rules prevents situations in which a
task sends particles to a neighbor and receives particles from another. The
task that receives particles from a neighbor computes the corresponding
non-bonding forces and sends back these quantities to the task that
actually owns the particles. Since the cost of the computation of non-bonding
forces grows (approximately) as the square of the number of particles
in each domain, the communication/computation ratio is favorable,
so that we obtain, even with relatively few particles, 
a reasonable efficiency, as confirmed by the results
reported in Table \ref{tab:dynlb}.

\begin{table}[h!]

\begin{centering}
\begin{tabular}{|c|c|c|}
\hline 
Number of tasks & Time (sec) & Efficiency (\%) \tabularnewline
\hline 
$32$&
230.4 &  \tabularnewline
$128$&
84.9 & 67.8 \tabularnewline
$512$&
27.9 & 51.6 \tabularnewline
$1024$& 14.2& 50.7 \\ \hline
\end{tabular}
\par\end{centering}

\caption{Times (in seconds) for 100 iterations of a 4000 particles biopolymer
translocation in a $128\times128\times128$ lattice.
Note that 32 is the minimum number of tasks with the
system configuration used for this test.
}

\label{tab:dynlb} 
\end{table}

This level of efficiency allows one to think of the possibilities
that lie ahead in this type of simulation, where
one will soon have at their
disposal IBM Blue-Gene systems with $\sim~10^5$ processors.
With this computational power, we estimate that it will be possible to simulate
$4\cdot10^4$ particles for $10^6$ time-steps
corresponding to approximately
$1\mu$sec over a region of about $1~\mu$m in one day.
% which translates to $\sim 0.1$ sec in a month. 
%The science implication to such an ability is profound.
One can then actually think of modelling atomistic simulation 
in the realm of nearly-macroscopic time-scales.
In such an approach, the break-even point for chemical 
specificity, where each bead would map one base-pair, could be reached,
marking the hand-shaking point with a new generation of multiscale
codes. This can be accomplished through using 
specific potentials in the code to account for the
molecular specificity of the different base-pairs.
However, the scientific problem, even with one base-pair per bead, of
deciding how to coarse grain the various atoms in the base-pair
will remain an open challenge.

\section{Numerical Set-up}

In our simulations we use a three-dimensional box 
of size $N_{x}\times N_{x}/2\times N_{x}/2$
in units of the lattice spacing $\Delta x$. The box contains both
the polymer and the fluid solvent. The former is initialized via a
standard self-avoiding random walk algorithm and further relaxed to
equilibrium by Molecular Dynamics. 
At time zero, the first bead of the polymer is placed at the
vicinity of the pore.
The solvent is initialized with
the equilibrium distribution corresponding to a constant density and
zero macroscopic speed. Periodicity is imposed for both the fluid
and the polymer in all directions. A separating wall is located in
the mid-section of the $x$ direction, at $x/\Delta x=N_{x}/2$, with
a square hole of side $h=3\Delta x$ at the center, through which
the polymer can translocate from one chamber to the other. For polymers
with up to $N_0=400$ beads we use $N_{x}=80$; for larger polymers
$N_{x}=100$. At $t=0$ the polymer resides entirely in the right
chamber at $x/\Delta x>N_{x}/2$. 

Translocation is induced by a constant electric force ($F_{drive}$)
which acts along the $x$ direction and is confined in a rectangular
channel of size $3\Delta x\times\Delta x\times\Delta x$ along the
streamwise ($x$ direction) and cross-flow ($y,z$ directions). The
solvent density and kinematic viscosity are $1$ and $0.1$, respectively,
and the temperature is $k_{B}T=10^{-4}$. All parameters are in units
of the LB timestep $\Delta t$ and lattice spacing $\Delta x$, which
we set equal to 1. Additional details have been presented in Ref.
\cite{ourLBM}. In our simulations we use $F_{drive}=0.02$ and a
friction coefficient $\gamma=0.1$. It should be kept in mind that
$\gamma$ is a parameter governing both the structural relation of
the polymer towards equilibrium and the strength of the coupling with
the surrounding fluid. With our parametrization, the process falls
in the fast translocation regime, where the total translocation time
is much smaller than the Zimm relaxation time. 

In order to interpret our results in terms of physical units, we map
the semiflexible polymers used in our simulations with the DNA typical
persistence length. With three lattice sites for a $12\, nm$ large
hole, the lattice site is $4\, nm$. With this mapping, our pore size
is close to the experimental pores which are of the order of $12$
nm and our MD particles correspond to about $30$ base pairs, 
reasonably close to the
persistence length of $\lambda$-phage double-strand DNA ($\sim 50\, nm$). 
The fact that the polymers modelled here have a persistence length
of about 10 monomers is also confirmed analytically by approaching
our polymers as worm like chains.
Moreover, the polymers presented here correspond
to DNA lengths in the range $0.8-20$ kbp whereas the DNA lengths used
in the experiments are larger (up to $\sim100$ kbp); the current
multiscale approach can be extended to handle these lengths, assuming
that appropriate computational resources are available.

\subsection{Translocation statistics}

\begin{figure}
\begin{centering}
\epsfig{file=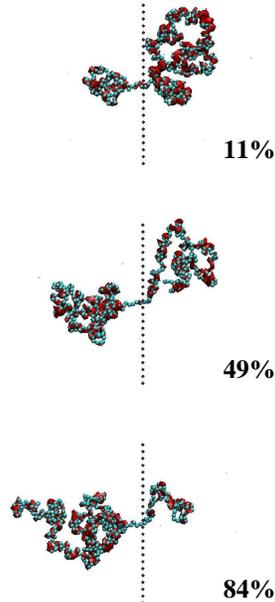,width=0.2\textwidth} 
\par\end{centering}
\caption{Snapshots of a typical event: a polymer 
($N_0=300$) translocating
from the right to the left is depicted at various stages of the process. The numbers
shown in each panel represent the fraction of the polymer
that has already passed through the pore.
The vertical dots in the middle of each panel represent the membrane wall.}
\label{FIG1} 
\end{figure}

Extensive simulations of a large number of translocation events over
$100-1000$ initial polymer configurations for each length confirm
that most of the time during the translocation process the polymer
assumes the form of two almost compact blobs on either side of the
wall: one of them (the untranslocated part, denoted by $U$) is contracting
and the other (the translocated part, denoted by $T$) is expanding.
Snapshots of a typical translocation event shown in Fig.~\ref{FIG1}
firmly support this picture. 

\begin{figure}
\begin{centering}
\includegraphics[width=0.3\textwidth]{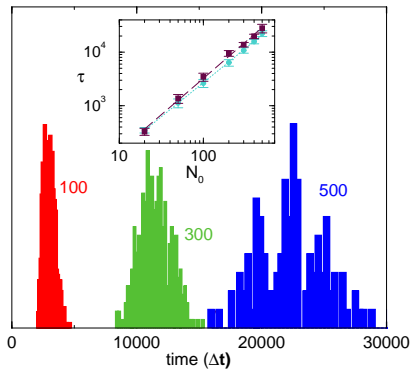} 
\par\end{centering}
\caption{Duration histograms for polymers of different
lengths. The distributions are skewed and the most probable
time from these distributions is used to construct the
scaling shown in the inset (cyan points). The scaling for
the case without a fluid solvent is also shown (black points).}
\label{FIG2} 
\end{figure}

The variety of different initial polymer realizations 
produce a scaling law dependence of the translocation times
on length \cite{Nelson}. 
We construct duration histograms by accumulating all events for each length. 
The resulting 
distributions deviate from simple gaussians and are 
skewed towards longer times (see Fig.~\ref{FIG2}). Hence, 
the translocation time for each length is not assigned to 
the mean, but to the most probable time, 
which is the position of the maximum in the histogram.
By calculating the most probable times
for each length, a superlinear relation between the
translocation time $\tau$ and the number of beads $N_0$ 
is obtained, as shown in Fig.~\ref{FIG2}(inset). The exponent
in the scaling law $\tau (N_0) \sim N_0^{\alpha}$ is calculated to be
$\alpha \sim 1.28\pm0.01$, for lengths up to $N_0=500$ beads.
The observed exponent is in very good agreement 
with a recent experiment on double-stranded DNA
translocation, that reported $\alpha\simeq1.27\pm0.03$ \cite{NANO}. 
This agreement makes it plausible that 
the generic polymers modeled in our simulations can be 
thought of as DNA molecules. Without hydrodynamics translocation
is significantly slowed down as indicated by a scaling
exponent $\alpha'  \sim 1.36\pm0.03$; data without the 
presence of a solvent are  also presented in Fig. \ref{FIG2}.

\subsection{Translocation dynamics}

\begin{figure}
\begin{centering}
\includegraphics[width=0.5\textwidth]{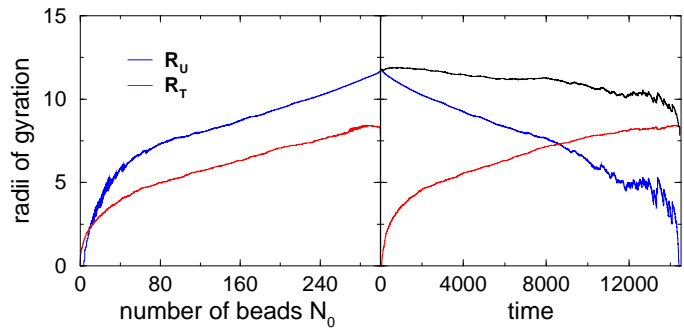} 
\par\end{centering}
\caption{Radii of gyration as functions of the number of beads (left panel) and
time (right panel) as an average over about 100 events for $N_0=300$.  Blue
is the untranslocated part $R_U$, red the translocated part $R_T$ and 
black the effective radius $R_{eff}$, defined in Eq. (\ref{R_eff}).}
\label{FIG3} 
\end{figure}

We next turn to the dynamics of the biopolymer as it passes
through the pore. A radius of gyration $R_{I}(t)$ (with
$I=U,T$) is assigned to each part of the polymer, the untranslocated
and translocated one; we can also define an effective radius 
through 
\begin{equation}
R_{eff} = \left[ R_T^{\nu} + R_U^{\nu}\right]^{1/\nu}
\label{R_eff}
\end{equation}
with $\nu \sim 0.6$, which applies to a self-avoiding random walk.
At the end of the translocation process, the radius of gyration is
considerably smaller than it was initially: $R_{T}(t_{X})<R_{U}(0)$,
where $t_{X}$ is the total translocation time for an individual event.
Taking averages over a few hundreds of events
for $N_0=200$ beads showed that $\lambda_{R}=R_{T}(t_{X})/R_{U}(0)\sim0.7$.
This reveals that as the polymer passes through the pore
it becomes more compact than it was at the initial stage of the event,
due to incomplete relaxation.
In Fig.~\ref{FIG3}, we represent both radii of gyration 
as functions of the number of beads and of the translocation time.
The curves shown in the figure are averages over about 100 events
for $N_0=300$. By definition, $R_U(t)$ vanishes
at $t=t_X$, while $R_T$ increases monotonically from 
$t=0$ up to $t=t_X$, although it never reaches the value $R_U(t=0)$.
The rates of change of the two radii are approximately equal and 
constant (see left panel of Fig. \ref{FIG3}), 
except at the end points of the event
where the radius of gyration itself is not a well defined quantity
because of the small number of beads included.
The effective radius $R_{eff}$ shows a very small decreasing slope, 
in other words, it is almost constant (right panel of Fig. \ref{FIG3}).

\begin{figure}
\begin{center}
\includegraphics[width=0.4\textwidth]{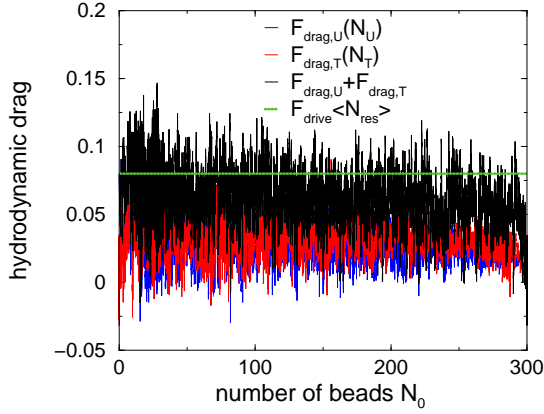} 
\caption{Viscous drag on the untranslocated and translocated parts 
of the polymer as a function of the number of beads and for an
average of 100 translocations events with $N_0=300$ beads.
The horizontal line denotes the electric drive in the pore region.}
\label{FIG4}
\end{center}
\end{figure}

Throughout its motion the polymer continuously interacts with the
fluid environment. The forces that essentially control the process
are the electric drive $\vec{F}_{drive,i}$ and the hydrodynamic
drag $\vec{F}_{drag,i}$ which act on each bead, for the untranslocated
($l=U$) and translocated ($l=T$) parts.. 
The drag forces $\vec{F}_{drag,i}$ on both parts of the polymer,
change in time. 
In Fig. \ref{FIG4}, both drag forces
are presented as a function
of time, together with their sum and the value of the electric drive in the 
pore, $F_{drive}=qE \langle N_r\rangle$, 
where $\langle N_r\rangle$ is the average number of 
resident monomers in the pore (simulations provide $N_r \sim 4.2$).
This figure, referring to an average over the entire ensemble
of trajectories for a $300$-monomer molecule, shows 
that the untranslocated strand alone can by no means
balance the external force; only the sum of the 
translocated and untranslocated drag forces comes close to balancing the drive.

Apart from these forces and specifically at the end points 
(initiation and completion of the passage through the
pore) entropic forces become important. The
fluctuations experienced by the polymer due to the presence of the
fluid are correlated to these entropic forces which, at 
least close to equilibrium, can be
expressed as the gradient of the free energy with respect to the
fraction of translocated beads. At the final stage of a
translocation event, the radius of the untranslocated part
undergoes a visible deceleration (see Fig.~\ref{FIG3}), the
majority of the beads having already translocated. It is, thus,
entropically more favorable to complete the passage through the
hole rather than reverting it, that is, the entropic forces
cooperate with the electric field and the translocation is
accelerated.

The entropic forces can also lead to rare events, such as
retraction, which occur in our simulations at a rate less than 2\%
and depend on length, initial polymer configuration and parameter
set. A retraction event is related to a polymer that
anti-translocates after having partially passed through the pore.
We have visually inspected the retraction events and associate
them with the translocated part entering a low-entropy
configuration (hairpin-like) subject to a strong entropic
pull-back force from the untranslocated part: The translocated
part of the polymer assumes an elongated conformation, which leads
to an increase of the entropic force from the coiled,
untranslocated part of the chain. As a result, the translocation
is delayed and eventually the polymer is retracted.

\subsection{Translocation work}

\begin{figure}
\begin{centering}
\includegraphics[width=0.5\textwidth]{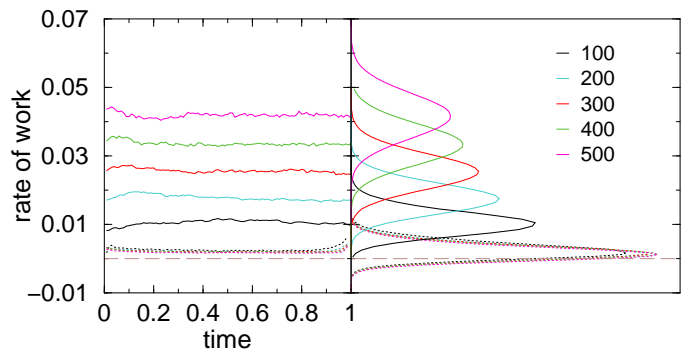} 
\par\end{centering}
\caption{Translocation work per unit time for the hydrodynamic 
(solid lines) and electric fields (dotted lines) for polymers
of different lengths. The right panel shows the probability
distributions of both work rates. The labels correspond to 
the number of monomers $N_0$ for each case.}
\label{FIG5} 
\end{figure}

As a final step, we study the work performed on the biopolymer
throughout its translocation. On general grounds, hydrodynamic
interactions are expected to minimize frictional effects and form
a cooperative background that assists the passage of the polymer
through the pore. We investigate the cooperativity of the
hydrodynamic field through 
the work ($W_{H}$) made by the fluid on the polymer per unit time:
\begin{equation}
\frac{dW_{H}}{dt}=\gamma \Big\langle\sum_i^{N_0}
\vec{v}_i(t) \cdot \vec{u}_i(t)\Big\rangle
\end{equation}
Through this
definition, positive values of this hydrodynamic work rate
indicate a cooperative effect of the solvent, while negative
values indicate a competitive effect by the solvent. The work $W_{E}$ done
per timestep by the electric field on the polymer can
also be easily obtained through the expression:
\begin{equation}
\frac{dW_E}{dt} =
 \Big\langle\sum_i^{N_0} \vec{F}_{drive,i} \cdot \vec{v}_{i}(t)\Big\rangle
\end{equation}
The brackets in the above equations denote
averages over different realizations of the polymer for the same
length. The results for the averages over all realizations are
qualitatively similar to the work rates for an individual event of
the same length. For all lengths studied here, we found that the
total work per timestep of the hydrodynamic field on 
the whole chain is essentially constant, as shown in Fig.~\ref{FIG5} 
for the averages over all events at each polymer length.
For all these cases, the electric work rates
are also constant with time. 
The hydrodynamic work per time is
larger than the corresponding electric field work, because
the latter only acts in the small region around the pore, 
and this is the reason why the electric field work is 
independent of polymer length. 

In addition to the variation of the work rates with time it is
useful to analyze their distributions during translocation events.
We show these in Fig. \ref{FIG5}, where it is evident that the
distribution of the hydrodynamic work rate 
lies entirely in the positive range,
indicating that hydrodynamics turns the solvent into a cooperative
environment, that is, it enhances the speed of the translocation process.
In the same figure, the distribution for the electric work rate
over all events
for all lengths considered is also shown; these distributions are mostly
positive but have a small negative tail which indicates that beads
can be found moving against the electric field.

\section{Conclusions \label{sec_concl} }

We have presented a new multiscale methodology based on the direct
coupling between atomic or molecular scale particle motion and mesoscopic hydrodynamics of
the surrounding solvent. Due to the particle-like nature of the
mesoscopic lattice Boltzmann solver, this coupling proceeds
seamlessly in space and time. Correlations between the atomic scale
and the hydrodynamic scale are also explicitly included through
direct and local interactions between the particles representing solvent motion
and the those belonging to the polymer. As a result, hydrodynamic interactions
between the polymer and the surrounding fluid are explicitly taken
into account, with no need of resorting to non-local
representations. This allows a state-of-the-art modeling of
biophysical phenomena, where hydrodynamic correlations play a
significant role. We have also described an efficient
parallel implementation of the method which exhibits excellent scalability
on the IBM BlueGene platform.

As an application we modelled the translocation of polymers, which
resemble DNA, through nanometer-sized pores.
Besides statistical properties, such as scaling exponents, the
present methodology affords direct insights into the details of
the {\it dynamics} as well as the {\it energetics} of the
translocation process, thereby offering a very valuable complement
to experimental investigations of these complex and fascinating
biological phenomena. It also shows a significant potential to
deal with problems that combine complex fluid motion and
dynamics at the molecular scale. 

\subsubsection*{Acknowledgments}

MF acknowledges support by Harvard's Nanoscale Science and Engineering
Center, funded by NSF (Award No. PHY-0117795). MB, SS and SM wish to thank
the Initiative for Innovative Computing of Harvard University for 
financial support and the Harvard Physics Department and School of 
Engineering and Applied Sciences for their hospitality.


\begin{thebibliography}{10}

\bibitem{MMS_review}
Lu, G. and Kaxiras, E.:
Overview of Multiscale Simulations of Materials, Handbook of Theoretical and Computational Nanothechnology , Vol. X:1-33, edited by M. Rieth and W. Schommers, American Scientific Publishers, 2005. 

\bibitem{LBEa}
Benzi, R. Succi, S., and Vergassola, M., Phys. Rep. 222:145-197,
1992.

\bibitem{LBEb}
D.A. Wolf--Gladrow, Lattice gas cellular automata and
lattice Boltzmann models, Springer Verlag, New York, 2000. 

\bibitem{LBEc}
Succi, S., The lattice Boltzmann equation. Oxford University Press, Oxford,
2001. 

\bibitem{ourLBM}Fyta, M.G., Melchionna, S., Kaxiras, E., and Succi,
S.: Multiscale coupling of molecular dynamics and hydrodynamics: application
to DNA translocation through a nanopore. Multiscale Model. Simul.
5:1156-1173, 2006. 

\bibitem{DUN}Ahlrichs, P. and Duenweg, B.: Lattice-Boltzmann simulation
of polymer-solvent systems. Int. J. Mod. Phys. C 9:1429-1438, 1999; 
Simulation of a single polymer chain in solution by combining
lattice Boltzmann and molecular dynamics. J. Chem. Phys. 111:8225-8239,
1999.

\bibitem{adhikari}Adhikari, R., Stratford, K., Cates, M.E. and Wagner, A.J.: 
Fluctuating Lattice-Boltzmann, Europhys.Lett. 71:473-477, 2005.

\bibitem{TRANSL}Lodish, H., Baltimore, D., Berk, A., Zipursky, S.,
Matsudaira, P., and Darnell, J.: Molecular Cell Biology, W.H. Freeman
and Company, New York, 1996.

\bibitem{EXPRM}Kasianowicz, J.J., et al: Characterization of individual
polynucleotide molecules using a membrane channel. Proc. Nat. Acad.
Sci. USA 93:13770-13773, 1996; Meller, A., et al: Rapid
nanopore discrimination between single polynucleotide molecules.
Proc. Nat. Acad. Sci. USA 
97:1079-1084, 2000; Li, J., et al: DNA molecules and configurations
in a solid-state nanopore microscope. Nature Mater. 2:611-615, 2003.

\bibitem{statisTrans}Sung, W. and Park, P.~J.: Polymer translocation
through a pore in a membrane. Phys. Rev. Lett. 77:783-786, 1996.

\bibitem{DynamPRL}Matysiak, S., et al: Dynamics of polymer translocation
through nanopores: Theory meets experiment. Phys. Rev. Lett. 96:118103,
2006.

\bibitem{Nelson}Lubensky, D.~K. and Nelson, D.~R.: Driven polymer
translocation through a narrow pore. Biophys. J. 77:1824-1838, 1999.

\bibitem{SHAKE} Ciccotti, G. and Ryckaert, J.-P.,: Molecular dynamics
simulation of rigid molecules. Comp. Phys. Rep. 4:345-392, 1986.

\bibitem{SMJCP}Melchionna, S.: Design of quasi-symplectic propagators
for Langevin dynamics. J. Chem. Phys., in press, 2007.

\bibitem{RATTLE} Andersen, H.C.: Rattle: a velocity version of the
SHAKE algorithm for molecular dynamics calculations. J. Comp.Phys.
52:24-34, 1983.

\bibitem{PARALB} Amati, G. Piva, R. and Succi, S.: Massively parallel
Lattice-Boltzmann simulation of turbulent channel flow.
Intl. J. Mod. Phys. C 4:869-877, 1997.

\bibitem{MD1} Boyer, L.L. Pawley, G.S.: Molecular dynamics
of clusters of particles interacting with pairwise forces using a
massively parallel computer. J. Comp. Phys. 78:405-423, 1988.

\bibitem{MD2} Heller, H. Grubmuller, H. Schulten, K.: Molecular
dynamics simulation on a parallel computer. Molec. Sim.
5:133-165, 1990.

\bibitem{MD3} Rapaport, D.: Multi-million particle molecular
dynamics: II.Design considerations for distributed processing.
Comp.Phys.Comm. 62:198-216, 1991.

\bibitem{MD4} Brown, D. Clarke, J.H.R. Okuda, M. Yamazaki, T.:
A domain decomposition parallelization strategy for molecular dynamics
simulations on distributed memory machines. Comp.Phys.Comm.
74:67-80, 1993. 

\bibitem{MD5} Esselink, K. Smit, B. Hilbers, P.A.J.: Efficient
parallel implementation of molecular dynamics on a toroidal network:
I.Parallelizing strategy. J. Comp. Phys. 106:101-107, 1993.

\bibitem{MD6} Plimpton, S.: Fast parallel algorithms for
short-range molecular dynamics. J. Comp. Phys. 117:1-19, 1995.

\bibitem{bluegene} \verb|http://domino.research.ibm.com/comm/research_projects.nsf/pages/bluegene.index.html|

\bibitem{NANO}Storm, A.~J. et al: Fast DNA translocation through
a solid-state nanopore. Nanolett. 5:1193-1197, 2005.


\end{thebibliography}
\end{document}